\documentstyle[12pt,a4,epsf]{article}

\newcommand{\ol}{\overline}
\newcommand{\wt}{\widetilde}
\def\diag{\mathop{\rm diag}\nolimits}

\begin{document}

\baselineskip=18pt plus 0.2pt minus 0.1pt

\begin{titlepage}
\title{\hfill\parbox{4cm}
       {\normalsize YITP-99-37\\{\tt hep-th/9906090}\\June 1999}\\
       \vspace{3.5cm}
       Branes in type 0/type II duality
       \vspace{2cm}}
\author{Yosuke Imamura\thanks{\tt imamura@yukawa.kyoto-u.ac.jp}
\\[7pt]
{\it Yukawa Institute for Theoretical Physics,}\\
{\it Kyoto University, Kyoto 606-8502, Japan}
}
\date{}

\maketitle
\thispagestyle{empty}

\vspace{1cm}

\begin{abstract}
\normalsize
We derive relations
between type 0 and type II D-brane configurations under the T-duality
suggested by Bergman and Gaberdiel
and confirm that the massless fields on D-branes are identical
to those on the dual D-brane configurations.
Furthermore, we discuss dualities of type 0 and type II NS5-branes
and find that the dual of an unwrapped type 0 NS5-brane is
a Kaluza-Klein monopole with non-supersymmetric blow up modes.
\end{abstract}


\vfill

\noindent

\end{titlepage}

\section{Introduction}

During the last year, type 0 string theories\cite{DH,SW},
which are superstring theories without spacetime supersymmetry,
have attracted much attention.
They allow us the possibility to analyze
non-supersymmetric QCD by means of the brane techniques
developed for supersymmetric theories\cite{wall}.
However, the properties of type 0 theories are not as well known as
other supersymmetric string theories,
because BPS conditions,
which provide a great deal of information concerning supersymmetric theories,
are not available.
Recently, Bergman and Gaberdiel found a duality
between type 0 and type II theories\cite{BG}.
With this duality, it may become possible to obtain
new properties of type 0 theories from knowledge
of type II theories.
Therefore, it is important to investigate this duality.
The purpose of this paper is to find
the duality relations between brane configurations
in type 0 theories and those in type II theories.

\section{Type 0/type II duality}
As is mentioned in \cite{BG}, type II theory
compactified on ${\bf S}^1$ with monodromy $(-1)^{F_s}$
is T-dual to type 0 theory on ${\bf S}^1$ with monodromy $(-1)^{F_R}$,
where $F_s$ and $F_R$ are operators that count spacetime fermions
and right-moving worldsheet fermions, respectively.%
\footnote{I would like to thank S. Sugimoto for clearing up
my misunderstanding of the definition of these operators.}
In Ref.\cite{BG}, these compactifications are represented as
compactifications on ${\bf S}^1/{\bf Z}_2$.
${\bf Z}_2$ is generated by an operator $S\cdot(-1)^{F_s}$ or $S\cdot(-1)^{F_R}$,
where $S$ is a half shift of ${\bf S}^1$.
These two expressions are the same,
except that the definitions of the radius of ${\bf S}^1$ are different by a factor $2$.
(Precisely, duals of type IIA and type IIB are type 0B and type 0A,
respectively.
We omit the letters A and B in what follows.)
Instead of explicitly showing the equivalence
of these two theories in terms of the world sheet CFT,
we only give the spectra of these theories in
Table \ref{typeii.tbl} and Table \ref{type0.tbl}.
In these tables, $n$ and $p^9$ represent the wrapping number
and the Kaluza-Klein momentum along the compactified direction.
\begin{table}[htb]
\caption{Spectrum of type II theory
on ${\bf S}^1$ with monodromy $(-)^{F_s}$.
The upper (lower) signs in the complex signs correspond to type IIB (IIA) theory.
The variables $n$ and $p^9$ represent the wrapping number
and the Kaluza-Klein momentum along the compactified direction.}
\label{typeii.tbl}
\begin{center}
\begin{tabular}{ccc}
\hline
\hline
                 &$p^9\in{\bf Z}/R$  &$p^9\in({\bf Z}+1/2)/R$ \\[1ex]
$n\in2{\bf Z}$   &(NS$+$,NS$+$)  &(NS$+$,R$+$) \\
                 &(R$\pm$,R$+$)    &(R$\pm$,NS$+$) \\[1ex]
$n\in2{\bf Z}+1$ &(NS$-$,NS$-$)  &(NS$-$,R$-$) \\
                 &(R$\mp$,R$-$)    &(R$\mp$,NS$-$)\\
\hline
\end{tabular}
\end{center}
\end{table}
\begin{table}[htb]
\caption{Spectrum of type 0 theory
on ${\bf S}^1$ with monodromy $(-)^{F_R}$.
The upper (lower) signs in the complex signs correspond to type 0B (0A) theory.
The variables $n$ and $p^9$ represent the wrapping number
and the Kaluza-Klein momentum along the compactified direction.}
\label{type0.tbl}
\begin{center}
\begin{tabular}{ccc}
\hline
\hline
                 &$p^9\in{\bf Z}/R$  &$p^9\in({\bf Z}+1/2)/R$ \\[1ex]
$n\in2{\bf Z}$   &(NS$+$,NS$+$)  &(NS$-$,NS$-$) \\
                 &(R$\pm$,R$+$)    &(R$\mp$,R$-$)  \\[1ex]
$n\in2{\bf Z}+1$ &(NS$+$,R$+$)   &(NS$-$,R$-$) \\
                 &(R$\pm$,NS$+$)   &(R$\mp$,NS$-$)\\
\hline
\end{tabular}
\end{center}
\end{table}
Indeed, according to the results given in these tables,
we find that these two spectra are identical if we swap the
wrapping number and the Kaluza-Klein momentum.
In particular, a type II (type 0) closed string
with wrapping number $1$ corresponds to
a type 0 (type II) Kaluza-Klein mode with momentum $p=1/(2R)$.
This implies the relations
\begin{equation}
2\pi R^{(II)}T^{(II)}=\frac{1}{2R^{(0)}},\quad
2\pi R^{(0)}T^{(0)}=\frac{1}{2R^{(II)}},
\end{equation}
where $T^{(II)}$ ($T^{(0)}$) and $R^{(II)}$ ($R^{(0)}$) are
the string tension and the compactification radius
in type II (type 0) theory.
Using these two equations, we obtain $T^{(II)}=T^{(0)}$ and
\begin{equation}
R^{(II)}R^{(0)}=\frac{l_s^2}{2},
\label{rr}
\end{equation}
where $l_s$ is the string length scale defined
with the common string tension
by $T^{(II)}=T^{(0)}=1/(2\pi l_s^2)$.
The relation (\ref{rr}) is different from that in usual
type II/type II T-duality
by the factor $1/2$ on the right-hand side.
This is the same as in the case of heterotic/heterotic T-duality where,
for example,
an $E_8\times E_8$ heterotic string with wrapping number $1$
is dual to a Kaluza-Klein mode with momentum $p=1/(2R)$
in the $({\bf16},{\bf16})$ representation of
unbroken $SO(16)\times SO(16)\in SO(32)$.

Concerning the string coupling constants $g^{(II)}$ and $g^{(0)}$,
we can obtain the following relation by requiring the
nine-dimensional Newton's constants of the two theories
to agree:
\begin{equation}
\frac{R^{(II)}}{(g^{(II)})^2}=\frac{R^{(0)}}{(g^{(0)})^2}.
\label{gg}
\end{equation}
\section{Dualities of D-branes}
Similarly to the type II/type II T-duality,
under the type 0/type II T-duality,
wrapped D-branes and unwrapped D-branes are
transformed into unwrapped D-branes and wrapped D-branes,
respectively.
However, the situation is more complicated
and interesting.

It is known that the tensions of type 0 D-branes are
$1/\sqrt2$ times those of type II D-branes\cite{KT}.
Namely, with the string coupling constant and the string length scale,
the tensions of D-branes are represented as
\begin{equation}
T_{Dp}^{(II)}=\frac{1}{(2\pi)^pl_s^{p+1}g^{(II)}},\quad
T_{Dp}^{(0)}=\frac{1}{\sqrt2}\frac{1}{(2\pi)^pl_s^{p+1}g^{(0)}}.
\label{tensions}
\end{equation}
Using (\ref{rr}), (\ref{gg}) and (\ref{tensions}),
we can obtain
\begin{equation}
T^{(II)}_{Dp}=2\times 2\pi R^{(0)}T^{(0)}_{D(p+1)}.
\end{equation}
This equation implies that an unwrapped type II D-brane is
transformed into a wrapped D-brane with wrapping number $2$ by the duality.
At first sight, this seems strange because
it seems that we can decompose the type 0 D-brane
into two wrapped D-branes with wrapping number $1$.
This is, however, impossible on the type II side.

\begin{figure}[htb]
\centerline{\epsfbox{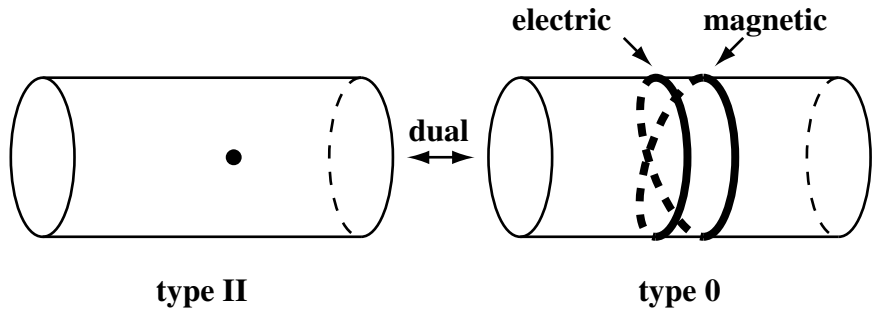}}
\caption{The T-dual of an unwrapped type II D$p$-brane is
a type 0 D($p+1$)-brane with wrapping number $2$.
Due to the monodromy $(-)^{F_R}$, an electric (magnetic) type 0 D-brane
changes into a magnetic (electric) D-brane
when it goes around ${\bf S}^1$.}
\label{dual.eps}
\end{figure}

This is interpreted as follows.
In the type 0 theory, by the monodromy $(-)^{F_R}$ around ${\bf S}^1$,
the two R-R fields $C$ and $\ol C$ are
transformed into $C$ and $-\ol C$, respectively.
(The fields $C$ and $\ol C$ are massless fields
in the (R$\pm$,R$+$) and (R$\mp$,R$-$) sectors, respectively.)
In terms of the D-brane charge, this implies that
when an electric (magnetic) D-brane goes around ${\bf S}^1$,
it is changed into a magnetic (electric) D-brane.
(We call D-branes `electric' when the signs of two R-R charges are the same
and `magnetic' when they are opposite,
although, except for D3-branes,
these are not electric nor magnetic in the usual meanings.)
Therefore, if the wrapping number of a D-brane is $1$,
it cannot be closed.
To connect two endpoints of a wrapped D-brane,
the wrapping number should be even.
This is the reason that a type 0 D-brane with wrapping number $2$,
which is dual to the single type II D-brane,
cannot be decomposed (Figure \ref{dual.eps}).

On a stack of $N$ type II D-branes, there exist $U(N)$ gauge bosons and
adjoint fermions,
and they make a vector multiplet of ten-dimensional ${\cal N}=1$
supersymmetry.
Following the argument above, the type 0 dual of this brane
configuration is a stack of $N$ electric and $N$ magnetic type 0 D-branes.
The field theory on the type 0 brane configuration is known to be
$U(N)\times U(N)$ Yang-Mills theory
with fermions belonging to the bi-fundamental representation\cite{dionic}.
However, now, because the electric D-branes and the magnetic D-branes are
connected, the gauge group is broken to the diagonal $U(N)$.
As a result, all massless fields on D-branes belong to the
adjoint representation of this $U(N)$, and
the massless spectrum agrees with that on type II D-branes.

\begin{figure}[htb]
\centerline{\epsfbox{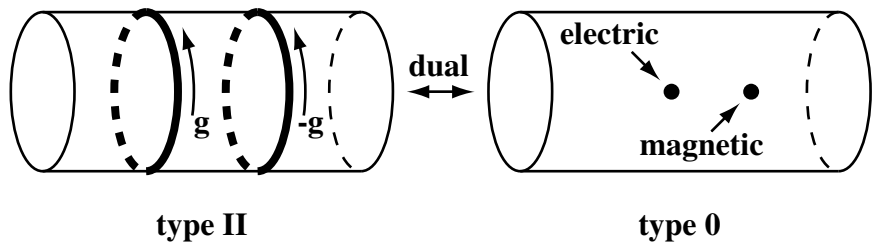}}
\caption{The charge of dual type 0 D-branes depends on the Wilson line
on the original wrapped type II D-branes.}
\label{dual2.eps}
\end{figure}

Next, let us consider the duality between
wrapped type II D-branes and unwrapped type 0 D-branes.
Using (\ref{rr}), (\ref{gg}) and (\ref{tensions})
we obtain
\begin{equation}
2\pi R^{(II)}T^{(II)}_{D(p+1)}=T^{(0)}_{Dp}.
\end{equation}
This equation implies that the dual of a wrapped type II D($p+1$)-brane
is an unwrapped type 0 D$p$-brane.
Is it electric or magnetic?
We should remember the fact that
because of the existence of the monodromy $(-)^{F_R}$,
we cannot globally define `electric' and `magnetic' D-branes.
Therefore, we can freely call a certain D-brane electric or magnetic.
However, it makes sense to say that two D-branes are of the same type or not.
If two type 0 D-branes are of different types,
where does the difference come from on the type II side?
To obtain an answer to this question,
let us consider an electric type 0 unwrapped D-brane.
It is dual to a type II wrapped D-brane.
On the type 0 side, we can change the D-brane
to a magnetic one by moving it around ${\bf S}^1$.
On the type II side, this corresponds to
a change of the Wilson line $g=\exp i\oint A_9dx^9$.
In the case of type II/type II duality,
to move an unwrapped D-brane round ${\bf S}^1$ corresponds
to a change of the Wilson line $g\rightarrow e^{2\pi i}g$.
Now, however, the size of ${\bf S}^1$ of type 0 theory
is $R^{(0)}=1/(2R^{(II)})$, unlike $1/R^{(II)}$ in type II/type II
duality,
and this implies that the change of the Wilson line should also be halved.
Namely, the change should be $g\rightarrow e^{i\pi}g$.
Therefore, we can conclude that
electric and magnetic unwrapped type 0 D-branes
correspond to wrapped type II D-branes
with Wilson lines $g$ and $-g$, respectively
(Fig.\ref{dual2.eps}).

We can show that the massless spectra on these type 0 and type II
D-branes agree.
Let us consider a stack of $N_1$ electric and $N_2$ magnetic unwrapped 
type 0 D-branes.
The massless fields on this stack are $U(N_1)\times U(N_2)$ gauge fields
and fermions in the $({\bf N}_1,\ol{\bf N}_2)$
and $(\ol{\bf N}_1,{\bf N}_2)$ representations.
The dual brane configuration is a stack of $N_1+N_2$ type II D-branes
with the Wilson line $\diag(g{\bf1}_{N_1},-g{\bf1}_{N_2})$,
where $g$ is a phase factor which does not affect the spectrum.
With the Wilson line, the gauge group $U(N_1+N_2)$ is
broken to $U(N_1)\times U(N_2)$.
Fermions in the adjoint representation are
decomposed into $({\bf adj}_1,{\bf1})$, $({\bf1},{\bf adj}_2)$,
$({\bf N}_1,\ol{\bf N}_2)$ and $(\ol{\bf N}_1,{\bf N}_2)$.
In addition to the Wilson line acting on these four representations as
$+1$, $+1$, $-1$ and $-1$,
we should take account of the monodromy $(-1)^{F_s}$,
which reverses the sign of wave functions of all the fermion fields.
As a result, fermions in
$({\bf N}_1,\ol{\bf N}_2)$ and $(\ol{\bf N}_1,{\bf N}_2)$
remain massless.
This is consistent with the type 0 spectrum.

\section{Dualities of NS5-branes}
As D-branes, NS5-branes play important roles
in brane constructions of Yang-Mills theories\cite{ak},
and they are important to establish the duality relations
of NS5-branes.
In this section, we give the duals of type 0 and type II
NS5-branes.

Regardless of the type of string theory in question,
we can obtain the tensions of NS5-branes
by using the Dirac quantization condition as follows:
\begin{equation}
T^{(0)}_{NS5}=\frac{1}{(2\pi)^5l_s^6(g^{(0)})^2},\quad
T^{(II)}_{NS5}=\frac{1}{(2\pi)^5l_s^6(g^{(II)})^2}.
\label{ns5s}
\end{equation}
From (\ref{gg}) and (\ref{ns5s}),
the relation
\begin{equation}
2\pi R^{(0)}T^{(0)}_{NS5}=2\pi R^{(II)}T^{(II)}_{NS5},
\end{equation}
is obtained.
This relation strongly suggests
that the dual of a wrapped type II NS5-brane is a wrapped type 0 NS5-brane.

In type II/type II T-duality, an unwrapped NS5-brane is transformed into
a Kaluza-Klein monopole with NUT charge $1$.
On the other hand, in the case of type 0/type II duality,
the dual of an unwrapped NS5-brane should be
a Kaluza-Klein monopole with NUT charge $2$.
The reason for this is as follows.
Because the momentum of Kaluza Klein modes along the compactified direction
is quantized as $n/(2R)$, the minimum electric charge coupled with the $U(1)$
gauge field $g_{\mu9}$ is $1/2$.
Therefore, due to the Dirac quantization condition,
the minimum NUT charge is $2$.

Kaluza-Klein monopoles with NUT charge $2$ can be made by
dividing the Taub-NUT manifold by ${\bf Z}_2$,
generated by a half shift of the ${\bf S}^1$ cycle.
Because there exists the monodromy $(-1)^{F_s}$ ($(-1)^{F_R}$)
in type II (type 0) theory,
the shift should be accompanied by the operator
$(-1)^{F_s}$ ($(-1)^{F_R}$).
The central region of this Kaluza-Klein monopole is ${\bf R}^4/{\bf Z}_2$.
${\bf Z}_2$ is generated by
${\cal P}(-1)^{F_s}$
(type II) or ${\cal P}(-1)^{F_R}$ (type 0), where $\cal P$ is
the parity operator which changes the coordinates $x^\mu$
on the ${\bf R}^4$ to $-x^\mu$.
Therefore, the origin of the manifold is the $A_1$ singularity.

Next, let us consider massless spectra on NS5-branes.
On every NS5-brane, there exist four massless scalar fields
representing the position of the brane.
In the dual picture, they appear as zero modes of NS-NS fields
on the Kaluza-Klein monopole.
In addition to these, on type II NS5-branes, a one-form $U(1)$ gauge field (type IIB)
or self-dual two-form and zero-form gauge fields (type IIA)
exist.\footnote
{We refer to scalar fields which do not correspond to the
fluctuation of branes as `zero-form gauge fields' because
they are magnetic duals of four-form gauge fields on five-branes.}
(Here, we focus only on bosonic fields.)
These gauge fields are represented as
zero modes of R-R fields on the Kaluza-Klein monopoles.
Indeed, there is one zero mode of a self-dual two-form field
on the Taub-NUT manifold.
Therefore, in type 0A theory, a zero mode of an R-R three-form
field $C_{\mu\nu i}$
(The indices $\mu$ and
$\nu$ label the directions contained in the Taub-NUT manifold,
and $i$ labels the flat directions parallel to the Kaluza-Klein
monopole.) gives a $U(1)$ gauge field, while
a zero mode of the other R-R three-form field
$\ol C_{\mu\nu i}$ is projected out by ${\bf Z}_2$.
In the same way, on the Kaluza-Klein monopole in type 0B theory,
there are zero modes of the four-form field $C_{\mu\nu ij}^+$
and the two-form field $C_{\mu\nu}$,
and they correspond to the self-dual two-form field and
zero-form gauge field on type IIA NS5-branes.
Zero modes of $\ol C_{\mu\nu ij}^-$ and $\ol C_{\mu\nu}$ are
projected out by ${\bf Z}_2$ again.

Conversely, by analyzing the type II Kaluza-Klein
monopoles, we can find gauge fields on type 0 NS5-branes.
On the type IIA Kaluza-Klein monopole,
there is a zero mode of the R-R three-form field $C_{\mu\nu i}$,
and on the type IIB Kaluza-Klein monopole,
zero modes of $C_{\mu\nu ij}^+$ and $C_{\mu\nu}$ exist.
Because the monodromy $(-1)^{F_s}$ acts on bosonic fields
trivially, none of these zero modes are projected out.

Furthermore, to obtain complete spectra,
we should consider the twisted sector at the
orbifold singularities.
From the (R$\mp$,R$-$) sectors of twisted closed strings at the
singularities, we obtain a one-form gauge field (type IIA)
or a zero-form gauge field and an anti-self-dual two-form field (type IIB).
Therefore, we conclude that
gauge fields on type 0 NS5-branes are doubled
and non-chiral, like the R-R fields in the bulk.
That is, in addition to four scalar fields
corresponding to fluctuations of branes,
there exist two zero-form gauge fields
and an unconstrained two-form field on a type 0A NS5-brane
and two one-form fields on a type 0B NS5-brane.\footnote
{In the previous version of this paper,
the argument regarding the twisted sector was missing.
The massless spectra on type 0 NS5-branes were
first obtained in Ref.\cite{NS5} using type 0A/type 0B duality.}
These fields are necessary for the
electric and magnetic D-branes to be attached independently
on NS5-branes.\cite{openp}

In the supersymmetric ${\bf R^4}/{\bf Z}_2$ orbifold case,
four massless scalar fields appear in the (NS$+$,NS$+$) sector of
twisted closed strings.
They belong to the $\bf(1,3)+(1,1)$ of the $SO(4)=SU(2)_L\times SU(2)_R$
rotation group and correspond to the (geometric and non-geometric)
blow-up modes of the singularity.
In the present case,
from the twisted (NS$-$,NS$-$) sector,
we have four massless scalar fields
belonging to the $\bf(3,1)+(1,1)$ representation.
Because they have chirality opposite to that of those in the supersymmetric
case, they correspond to `non-supersymmetric blow up modes';
they express a blow up respecting hyper-K\"ahler 
structure that differs from that of the Kaluza-Klein monopole before the
blow up.
If these modes are switched on, all the supersymmetries are broken.
Due to this property,
they can acquire mass, and, perhaps, they are removed from the massless spectrum.
After the blow up, the manifold becomes smooth and simply connected.
Therefore the wrapping number of a string is not conserved and,
at first sight, it seems that the GSO projection,
which depends on the wrapping number, is ill-defined.
However, this is not a problem, as explained below.
Because the worldsheet fermions $\psi^\mu$ (We use NS-R
formalism.) take values in the tangent bundle of the target space
at $X^\mu$, the boundary conditions of $\psi^\mu$
depend on the target space holonomy along closed strings.
Using the complex coordinates $\Psi^1=\psi^6+i\psi^7$,
$\Psi^2=\psi^8+i\psi^9$ and their hermitian conjugates
$(\Psi^i)^\dagger$ ($i=1,2$),
the action of $SU(2)_L\times SU(2)_R$ rotation is expressed as
\begin{equation}
U\rightarrow g_LUg_R,\quad
U\equiv
\left(\begin{array}{cc}
      \Psi^1 & -(\Psi^2)^\dagger \\
      \Psi^2 & (\Psi^1)^\dagger
\end{array}\right),\quad
g_L\in SU(2)_L,\quad
g_R\in SU(2)_R.
\end{equation}
The Kaluza-Klein monopole with non-supersymmetric
blow up is approximately divided into two parts:
the central region, which is almost $A_1$ ALE,
and the outer region, which is only weakly affected by the blow up modes.
These two regions possess different hyper-K\"ahler structures,
and holonomies of cycles in the outer and central regions
are elements of $SU(2)_L$ and $SU(2)_R$, respectively.
Let us consider a process in which
a wrapped string in the outer region moves adiabatically
into the central region and the wrapping becomes loose.
In the initial and final string configurations,
the holonomy along the string is ${\bf1}_2$ (the rank two unit matrix).
On a string passing across the border between the two regions,
where the target space is approximately ${\bf R}^4/{\bf Z}_2$,
the holonomy is $-{\bf1}_2$.
Connecting these with passes in $SU(2)_L$ and $SU(2)_R$,
we obtain the change of the holonomy during
the adiabatic process as follows:
\begin{equation}
\begin{array}{ccccccccc}
\makebox[0em][c]{the outer region}&&&&
\makebox[0em][c]{the border}&&&&
\makebox[0em][c]{the central region}
\\
{\bf1}_2
& \rightarrow & g_L\in SU(2)_L
& \rightarrow & -{\bf1}_2
& \rightarrow & g_R\in SU(2)_R
& \rightarrow & {\bf1}_2
\end{array}
\end{equation}
More explicitly, for example,
we can choose the path of the string so that
$g_L=\exp(i\theta_1\sigma_z)$ and $g_R=\exp(i\theta_2\sigma_z)$,
where $\theta_1$ ($\theta_2$) is a parameter
in the outer (central) region changing from $0$ to $\pi$ (from $\pi$ to $0$).
In this case, the phases $\omega_i$ in the boundary conditions
of the fermions $\Psi^i(\sigma+2\pi)=-\exp(i\omega_i)\Psi^i(\sigma)$ are
change as
\begin{equation}
\begin{array}{ccccccccccc}
&&
\makebox[0em][c]{the outer region}&&&&
\makebox[0em][c]{the border}&&&&
\makebox[0em][c]{the central region}
\\
\omega_1 & :  & 0
& \rightarrow & \theta_1
& \rightarrow & \pi
& \rightarrow & \theta_2
& \rightarrow & 0
\\
\omega_2 & :  & 0
& \rightarrow & -\theta_1
& \rightarrow & -\pi
& \rightarrow & \theta_2-2\pi
& \rightarrow & -2\pi.
\end{array}
\end{equation}
(Here, we are considering the NS-NS sector.
For the R-R sector, we should use the boundary conditions
$\Psi^i(\sigma+2\pi)=+\exp(i\omega_i)\Psi^i(\sigma)$.)
Therefore, this process causes a shift of the oscillators
of $\Psi^2$
($\Psi^2_{n+1/2}\rightarrow\Psi^2_{n-1/2}$ for the left-moving fermions
and $\wt\Psi^2_{n-1/2}\rightarrow\wt\Psi^2_{n+1/2}$ for the right-moving fermions)
while the oscillators of $\Psi^1$ do not change
($\Psi^1_{n+1/2}\rightarrow\Psi^1_{n+1/2}$
and $\wt\Psi^1_{n+1/2}\rightarrow\wt\Psi^1_{n+1/2}$).
This shift implies that even if we started from a string
in the ground state
$|0\rangle\otimes|0\rangle$, we have an excited state
$\Psi_{-1/2}^2|0\rangle\otimes(\wt\Psi_{1/2}^2)^\dagger|0\rangle$
after the adiabatic process.
These two states have different eigenvalues of operators $(-1)^{F_R}$
and $(-1)^{F_L}$,
where $F_L$ is operator that counts
left-moving worldsheet fermions.
If the initial state is (not) projected away by the GSO projection,
the final state also should (not) be projected away.
Therefore, we should impose different GSO projection
on the initial and final states.

Finally, I would like to note that
the scenario described above does not work for type 0 Kaluza-Klein monopoles,
which are dual to type II NS5-branes, because
the $(-)^{F_R}$ twist cannot be explained by holonomies of the target space.
This is consistent with the fact that
no bosonic fields appear in the twisted sector of type 0 strings.
\section{Conclusions}
We have derived the following duality relations
between type II and type 0 brane configurations.
\begin{itemize}
\item
The dual of an unwrapped type II D$p$-brane is
a stack consisting of a wrapped type 0 electric D($p+1$) brane
and a wrapped type 0 magnetic D($p+1$) brane.
\item
The dual of a wrapped type II D($p+1$)-brane is
an unwrapped type 0 D-brane.
The charge of the type 0 D-brane depends on the
Wilson line on the type II D-brane.
\item
The dual of a type 0 (type II) wrapped NS5-brane is
a type II (type 0) wrapped NS5-brane.
\item
The dual of a type 0 (type II) unwrapped NS5-brane is
a Kaluza-Klein monopole with NUT charge $2$ ($A_1$-singularity)
in type II (type 0) theory.
At the singularity in type II theory,
there exist non-supersymmetric blow up modes,
which probably acquire mass and are removed from the
massless spectrum.
\end{itemize}
\section*{Acknowledgements}
This work is supported in part by a Grant-in-Aid for Scientific
Research from the Ministry of Education, Science, Sports and Culture
(\#9110).

\end{document}